\documentclass[a4paper]{jpconf}
\usepackage{graphicx}
\usepackage{amssymb}
\usepackage{amsmath}
\usepackage{amsthm}
\usepackage{latexsym}

\def\LL{{\cal L}}
\newcommand{\radiustar} {{\rm R}_{\rm S}}

\begin{document}
\title{A perturbative approach for the study of compatibility
between nonminimally coupled gravity and Solar System experiments}

\author{Orfeu Bertolami$^1$, Riccardo March$^2$ and Jorge P\'aramos$^3$}

\address{$^1 $Departamento de F\'{\i}sica e Astronomia, Faculdade de Ci\^encias, Universidade do Porto,\\
Rua do Campo Alegre 687, 4169-007 Porto, Portugal\\
$^2 $Istituto per le Applicazioni del Calcolo, CNR, Via dei Taurini 19, 00185 Roma, Italy,\\
and INFN - Laboratori Nazionali di Frascati (LNF), Via E. Fermi 40 Frascati, 00044 Roma, Italy\\
$^3 $Instituto de Plasmas e Fus\~ao Nuclear, Instituto Superior T\'ecnico\\
Av. Rovisco Pais 1, 1049-001 Lisboa, Portugal}

\ead{orfeu.bertolami@fc.up.pt,r.march@iac.cnr.it,paramos@ist.edu}

\begin{abstract}
We develop a framework for constraining a certain class of theories of nonminimally coupled (NMC) gravity with Solar System observations.
\end{abstract}

\section{Introduction}
We consider the possibility of constraining a class of theories of nonminimally coupled gravity \cite{BBHL}
by means of Solar System experiments.
NMC gravity is an extension of $f(R)$ gravity where the action integral of General Relativity (GR) is modified in such
a way to contain two functions $f^1(R)$ and $f^2(R)$ of the space-time curvature $R$. The function $f^1(R)$ has a
role analogous to $f(R)$ gravity, and the function $f^2(R)$ yields a nonminimal coupling between curvature and
the matter Lagrangian density.
For other NMC gravity theories and their potential applications, see, e.g., \cite{PO1,PO2,PO3,Io1,Io2}.

NMC gravity has been applied to several astrophysical and cosmological problems such as
dark matter \cite{dm1BP,dm2BFP}, cosmological perturbations \cite{pertBFP}, post-inflationary reheating \cite{reheating} or the current accelerated expansion of the Universe \cite{BFP}.

In the present communication, by extending the perturbative study of $f(R)$ gravity in \cite{CSE},
we discuss how a general framework for the study of Solar System constraints to NMC gravity can be developed. The approach is based on a suitable linearization of the field equations of NMC gravity
around a cosmological background space-time, where the Sun is considered as a perturbation. Solar System observables
are computed, then we apply the perturbative approach to the NMC model by
Bertolami, Fraz\~ ao and P\'aramos \cite{BFP}, which constitutes a natural
extension of $1\slash R^n$ ($n>0$) gravity \cite{CDTT} to the non-minimally coupled case.
Such a NMC gravity model is able to predict the observed accelerated expansion of the Universe.
We show that, differently from the pure $1\slash R^n$ gravity case, the NMC model cannot be constrained
by this perturbative method so that it remains, in this respect, a viable theory of gravity.
Further details about the subject of the present communication can be found in the manuscript \cite{BMP}.

\section{NMC gravity model}\label{sec:NMC-model}

We consider a gravity model with an action functional of the type \cite{BBHL},
\[
S = \int \left[\frac{1}{2}f^1(R) + [1 +  f^2(R)] \LL_m \right]\sqrt{-g} \, d^4x,
\]
where $f^i(R)$ ($i=1,2$) are functions of the Ricci scalar curvature $R$, $\LL_m$ is the Lagrangian
density of matter and $g$ is the metric determinant. By varying the action with respect to the metric we get the field equations
\begin{equation}\label{field-eqs}
  \left(f^1_R + 2 f^2_R \LL_m \right) R_{\mu\nu} - \frac{1}{2} f^1 g_{\mu\nu} =  \left(1 +  f^2 \right) T_{\mu\nu} +
\nabla_{\mu\nu} \left(f^1_R + 2 f^2_R \LL_m \right),
\end{equation}
where $f^i_R \equiv df^i\slash dR$ and $\nabla_{\mu\nu} = \nabla_\mu \nabla_\nu -g_{\mu\nu}\square$. We describe matter as a perfect fluid with negligible pressure:
the Lagrangian density of matter is $\LL_m = -\rho$ and the trace of the energy-momentum tensor is $T = -\rho$.
We write $\rho = \rho^{\rm cos} + \rho^{\rm s}$, where $\rho^{\rm cos}$ is the cosmological mass density
and $\rho^{\rm s}$ is the Sun mass density.

We assume that the metric which describes the spacetime around the Sun is a perturbation of a flat Friedmann-Robertson-Walker (FRW) metric with scale factor $a(t)$:
\[
ds^2 = -\left[1 + 2\Psi(r,t) \right] dt^2 + a^2(t)\left(\left[1 + 2\Phi(r,t)\right] dr^2
+ r^2 d\Omega^2 \right),
\]
where $|\Psi(r,t)| \ll 1$ and $|\Phi(r,t)| \ll 1$.
The Ricci curvature of the perturbed spacetime is expressed as the sum
\[
R(r,t) = R_0(t) + R_1(r,t),
\]
where $R_0$ denotes the scalar curvature of the background FRW spacetime and $R_1$ is the perturbation due to
the Sun. Following Ref. \cite{CSE}, we linearize the field equations assuming that
\begin{equation}\label{R1-cond}
\left\vert R_1(r,t) \right\vert \ll R_0(t),
\end{equation}
both around and inside the Sun. This assumption means that the curvature $R$ of the perturbed spacetime remains close to the cosmological value $R_0$ inside the Sun. In GR such a property of the curvature is not satisfied inside the Sun. However, for $f(R)$ theories which are characterized by a small value of a suitable mass parameter (see next section),
condition (\ref{R1-cond}) can be satisfied. For instance, the $1\slash R^n$ ($n>0$) gravity model \cite{CDTT} satisfies condition (\ref{R1-cond}), as shown in \cite{CSE,HMV}.

Eventually, we assume that functions $f^1(R)$ and $f^2(R)$ admit a
Taylor expansion around $R=R_0$ and that terms nonlinear in $R_1$ can be neglected in the expansion.
We use the notation introduced by \cite{CSE} (for $i=1,2$):
\[
f^i_0 \equiv f^i(R_0)~~,~~  f^i_{R0}  \equiv \frac{d f^i}{dR}(R_0)~~,~~ f^i_{RR0}  \equiv \frac{d^2 f^i}{dR^2}(R_0).
\]

\section{Solution of the linearized field equations}

The details of the following computations can be found in the paper \cite{BMP}.
First we linearize the trace of the field equations (\ref{field-eqs}). Using condition (\ref{R1-cond}),
we neglect $O(R_1^2)$ contributions but we keep the cross-term $R_0 R_1$. Introducing the potential
$U = \left( f^1_{RR0} + 2 f^2_{RR0}\LL_m\right) R_1$, we get
\[
 \nabla^2 U -m^2 U = -\frac{1}{3}\left( 1 +  f^2_0 \right) \rho^{\rm s} +
\frac{2}{3} f^2_{R0}\rho^{\rm s}R_0 + 2 \rho^{\rm s} \square f^2_{R0} +
2 f^2_{R0} \nabla^2 \rho^{\rm s},
\]
where $m^2$ denotes the mass parameter
\begin{equation}\label{mass-formula}
m^2(r,t) =  \frac{1}{3}\bigg[\frac{f^1_{R0} -  f^2_{R0}\LL_m}{f^1_{RR0} + 2 f^2_{RR0}\LL_m}
- R_0 -   \frac{3 \square\left( f^1_{RR0} - 2 f^2_{RR0}\rho^{\rm cos}\right)
- 6\rho^{\rm s} \square f^2_{RR0}}{f^1_{RR0} + 2 f^2_{RR0}\LL_m} \bigg].
\end{equation}
When $f^2(R)=0$ we recover the mass formula of $f(R)$ gravity theory found in \cite{CSE}.
In the following we assume that $|mr| \ll 1$ at Solar System scale. Under this assumption
the solution for $R_1$ outside the Sun is given by
\begin{equation}\label{R1-solution}
R_1(r,t) = \left[\frac{-\frac{1}{3}\left( 1 + f^2_0 \right) + \frac{2}{3} f^2_{R0}R_0 +
2 \square f^2_{R0}}{ 4\pi \left( 2 f^2_{RR0}\rho^{\rm cos}-  f^1_{RR0}\right)}\right] \frac{M_{\rm S}}{ r},
\end{equation}
where $M_{\rm S}$ is the mass of the Sun. Then we linearize the field equations (\ref{field-eqs}) obtaining
\begin{eqnarray*}
&& \left( f^1_{R0} + 2 f^2_{R0}\LL_m \right) \left( \nabla^2\Psi + \frac{1}{2}R_1 \right) -  \nabla^2 \left[ \left( f^1_{RR0} + 2 f^2_{RR0}\LL_m \right) R_1 \right] = \left( 1 +  f^2_0 \right) \rho^{\rm s} - 2 f^2_{R0} \nabla^2 \rho^{\rm s},\\ \nonumber
&&\left( f^1_{R0} + 2 f^2_{R0}\LL_m \right)\left( -\frac{d^2\Psi}{dr^2} + \frac{2}{r}\frac{d\Phi}{dr} \right) -  \frac{1}{2}f^1_{R0}R_1 + \frac{2}{r}f^1_{RR0}\frac{dR_1}{dr} + \frac{4}{r} f^2_{RR0}\frac{\partial\left(\LL_m R_1\right)}{\partial r}
= \frac{4}{r} f^2_{R0}\frac{d\rho^{\rm s}}{dr}.
\end{eqnarray*}
Using the divergence theorem and the solution (\ref{R1-solution}) for $R_1$, from the first equation
we obtain the function $\Psi$ outside of the Sun:
\[
\Psi(r,t) = -\frac{2}{3r}\left( 1 +  f^2_0 +  f^2_{R0}R_0 + 3\square f^2_{R0} \right)
\int_0^{\radiustar} \frac{\rho^{\rm s}(x)}{f^1_{R0} + 2 f^2_{R0}\LL_m(x)}r^2\, dr,
\]
where $\radiustar$ is the radius of the Sun. If the following condition is satisfied,
\begin{equation}\label{Newt-limit}
\left\vert 2 f^2_{R0} \right\vert \rho^{\rm s}(r) \ll
\left\vert f^1_{R0} - 2 f^2_{R0}\rho^{\rm cos}(t) \right\vert, \qquad r \leq \radiustar,
\end{equation}
then the function $\Psi$ is a Newtonian potential:
\begin{equation}\label{Psi-solution}
\Psi(r,t) =  -\frac{GM_{\rm S}}{r},
\qquad G(t) = \frac{ 1 +  f^2_0 +  f^2_{R0}R_0 + 3\square f^2_{R0} }{6\pi \left( f^1_{R0} - 2 f^2_{R0}\rho^{\rm cos} \right)}
\qquad r \geq \radiustar,
\end{equation}
where $G(t)$ is an effective gravitational constant. Since $G$ depends on slowly varying cosmological quantities
we have $G(t) \simeq {\rm constant}$, so that $\Psi(r,t) \simeq \Psi(r)$.

The solution for the function $\Phi$ is computed from the second of the linearized field equations,
and we obtain $\Phi(r) = - \gamma \Psi(r)$, where the PPN parameter $\gamma$ depends on cosmological quantities and it is  given by
\[
\gamma = \frac{1}{2} \, \left[\frac{1 +  f^2_0 + 4 f^2_{R0}R_0 + 12 \square f^2_{R0}}
{1 +  f^2_0 +  f^2_{R0}R_0 + 3\square f^2_{R0}} \right].
\]
When $f^2(R)=0$ we find the known result $\gamma = 1\slash 2$ which holds for $f(R)$ gravity theories which
satisfy the condition $|mr| \ll 1$ and condition $\left\vert R_1 \right\vert \ll R_0$, as it has been shown in \cite{CSE}.
The $1\slash R^n$ ($n>0$) gravity theory \cite{CDTT}, where $f(R)$ is proportional to
$\left( R + {\rm constant}\slash R^n \right)$, is one of such theories that, consequently, have to be ruled out
by Cassini measurement.

\section{Application to a NMC cosmological model}

We consider the NMC gravity model proposed in \cite{BFP} to account for the observed accelerated expansion of the Universe:
\begin{equation}\label{case-study}
f^1(R) = 2\kappa R, \qquad f^2(R) = \left( \frac{R}{R_n} \right)^{-n}, \quad n>0,
\end{equation}
where $\kappa = c^4/16\pi G_N$, $G_N$ is Newton's gravitational constant, and $R_n$ is a constant.
This model yields a cosmological solution with a negative deceleration parameter $q <0$, and the scale factor $a(t)$ of the background metric follows the temporal evolution $a(t) = a_0 \left( t \slash t_0 \right)^{2(1+n)/3}$, where $t_0$ is the current age of the Universe.
Using the properties of the cosmological solution found in \cite{BFP} the mass parameter (\ref{mass-formula}) can be computed obtaining (we refer to \cite{BMP} for details of the computation):
\[
m^2 = \frac{ \mu(n)\rho^{\rm cos}+ \nu(n)\rho^{\rm s}}{\rho^{\rm cos}+\rho^{\rm s}} R_0, \qquad
R_0(t) = \frac{4(1+4n) (1+n) }{3t^2},
\]
where $\mu(n)$ and $\nu(n)$ are rational functions of the exponent $n$.
In \cite{BMP} it is shown that the condition $|mr| \ll 1$ imposes the extremely mild constraint $n \gg (1\slash 6){\rm R}_{S}^2 R_0 \sim 10^{-25}$.
Moreover, from the properties of the cosmological solution
\cite{BFP} we have $f^2_{R0}\rho^{\rm cos}(t)/\kappa = -2n \slash (4n + 1)$, from which it follows that condition
(\ref{Newt-limit}) is incompatible with the previous constraint $n \gg 10^{-25}$:
\[
 \left\vert \frac{\kappa}{f^2_{R0}\rho^{\rm cos}(t)} - 1 \right\vert \rho^{\rm cos} = \left( 3 + \frac{1}{2n} \right) \rho^{\rm cos}(t) \gg \rho^{\rm s}(r) \rightarrow  n \ll \frac{\rho^{\rm cos}}{2\rho^{\rm s}} \sim 10^{-33}.
\]
We now check the assumption $\left\vert R_1 \right\vert \ll R_0$.
The previous result shows that we can not rely on the validity of Newtonian approximation. Hence
we cannot use the effective gravitational constant $G$ defined in (\ref{Psi-solution}) for the estimate of the ratio $R_1 \slash R_0$, so that we resort to Newton's gravitational constant $G_N = c^4/16\pi\kappa$. The value of this ratio outside the Sun can be computed from the exterior solution (\ref{R1-solution}) for $R_1$, while the result for the interior solution requires a more involved computation, based on a polynomial model of the mass density $\rho^{\rm s}$, that can be found in \cite{BMP}:
\[
\frac{R_1}{R_0} \approx \frac{1+4n}{n (1+n)}\,\frac{G_NM_{\rm S} }{r} \quad\mbox{for } r \geq \radiustar, \qquad
\frac{R_1}{R_0} \approx \frac{1}{1+n} \quad\mbox{for } r < \radiustar.
\]
Though $\left\vert R_1 \right\vert \ll R_0$ for $n \gg 1$, the interior solution shows that non-linear terms in the Taylor expansion of $f^2(R)$ cannot be neglected, contradicting our assumption at the end of Section \ref{sec:NMC-model}:
\[
f^2(R) = f_0^2\bigg[ 1 - n \frac{ R_1}{R_0} + \frac{n(n+1)}{2} \left(\frac{R_1}{R_0}\right)^2  - \frac{1}{6}n(n+1)(n+2) \left(\frac{R_1}{R_0}\right)^3 \bigg] + O\left( \left( \frac{R_1}{R_0} \right)^4 \right).
\]
The lack of validity of the perturbative regime leads us to conclude that the model (\ref{case-study}) cannot be constrained by this method, so that it remains, in this respect, a viable theory of gravity.


\section*{References}
\begin{thebibliography}{9}
\bibitem{BBHL} Bertolami O, B\"{o}hmer C G, Harko T and Lobo F S N 2007 {\it Phys. Rev. D} {\bf 75} 104016
\bibitem{PO1} Puetzfeld D and Obukhov Y N 2013 {\it Phys. Rev. D} {\bf 87} 044045
\bibitem{PO2} Puetzfeld D and Obukhov Y N 2013 {\it Phys Lett A} {\bf 377} 2447
\bibitem{PO3} Puetzfeld D and Obukhov Y N 2013 Equations of motion in gravity theories with nonminimal coupling: a
loophole to detect torsion macroscopically? {\it Preprint} arXiv:1308.3369
\bibitem{Io1} Iorio L 2013 A Closer Earth and the Faint Young Sun Paradox: Modification of the Laws of Gravitation, or
Sun/Earth Mass Losses? {\it Preprint} arXiv:1306.3166
\bibitem{Io2} Iorio L 2013 Orbital effects induced by a certain class of modified theories of gravity with nonminimal
coupling between the matter and the metric {\it Preprint} arXiv:1306.3886
\bibitem{dm1BP} Bertolami O and P\'aramos J 2010 {\it JCAP} {\bf 03} 009
\bibitem{dm2BFP} Bertolami O, Fraz\~ao P and P\'aramos J 2012 {\it Phys. Rev. D} {\bf 86} 044034
\bibitem{pertBFP} Bertolami O, Fraz\~ ao P and P\'aramos J 2013 {\it JCAP} {\bf 05} 029
\bibitem{reheating} Bertolami O, Fraz\~ao P and P\'aramos J 2011 {\it Phys. Rev. D} {\bf 83} 044010
\bibitem{BFP} Bertolami O, Fraz\~ ao P and P\'aramos J 2010 {\it Phys. Rev. D} {\bf 81} 104046
\bibitem{CSE} Chiba T, Smith T L and Erickcek A L 2007 {\it Phys. Rev. D} {\bf 75} 124014
\bibitem{CDTT} Carroll S M, Duvvuri V, Trodden M and Turner M S 2004 {\it Phys. Rev. D} {\bf 70} 043528
\bibitem{BMP} Bertolami O, March R and P\'aramos J 2013 Solar System constraints to nonminimally coupled gravity {\it Preprint} arXiv:1306.1176
\bibitem{HMV} Henttunen K, Multam\" aki T and Vilja I 2008 {\it Phys. Rev. D} {\bf 77} 024040
\end{thebibliography}
\end{document}